\documentclass[a4paper]{jpconf}
\usepackage[english]{babel}
\usepackage[T1]{fontenc}
\usepackage{graphicx}
\usepackage{subfigure}
\usepackage{amssymb,amsmath,bm}
\newcommand{\goodgap}{%
	\hspace{\subfigtopskip}%
	\hspace{\subfigbottomskip}}
\usepackage[colorlinks=true,linkcolor=blue,citecolor=blue,urlcolor=black]{hyperref}
\usepackage{indentfirst}

\begin{document}
\title{From hypernuclei to the Inner Core of Neutron Stars: A Quantum Monte Carlo Study}

\author{D Lonardoni$^1$, F Pederiva$^{2,3}$, S Gandolfi$^4$}
\address{$^1$Physics Division, Argonne National Laboratory, Argonne, IL 60439, USA}
\address{$^2$Dipartimento di Fisica, Universit\`a di Trento, via Sommarive 14, I-38123 Trento, Italy}
\address{$^3$I.N.F.N. - T.I.F.P.A, Trento Institute for Fundamental Physics and Applications, I-38123 Trento, Italy}
\address{$^4$Theoretical Division, Los Alamos National Laboratory, Los Alamos, New Mexico 87545, USA}

\ead{lonardoni@science.unitn.it, pederiva@science.unitn.it, stefano@lanl.gov}

\begin{abstract}
Auxiliary Field Diffusion Monte Carlo (AFDMC) calculations have been employed to revise the interaction beween $\Lambda$-hyperons and nucleons in hypernuclei. The scheme used to describe the interaction, inspired by the phenomenological Argonne-Urbana forces, is the $\Lambda N+\Lambda NN$ potential firstly introduced by Bodmer, Usmani~\emph{et al.}. Within this framework, we performed calculations on light and medium mass hypernuclei in order to assess the extent of the repulsive contribution of the three-body part. By tuning this contribution in order to reproduce the $\Lambda$ separation energy in $^5_\Lambda$He and $^{17}_{~\Lambda}$O, experimental findings are reproduced over a wide range of masses. Calculations have then been extended to $\Lambda$-neutron matter in order to derive an analogous of the symmetry energy to be used in determining the equation of state of matter in the typical conditions found in the inner core of neutron stars. 
\end{abstract}

\section{Introduction}
The composition of the inner core of neutron stars (NS) still remains a largely unsolved question. The fairly recent observation of neutron stars with masses of order $2M_\odot$~\cite{Demorest:2010,Antoniadis:2013} has set a rather strong constraint on the stiffness of the equation of state (EoS) of hadronic matter at high densities. On the other hand simple physical arguments can be made, that introduce mechanisms for softening the EoS at high densities. One of them is based on the idea that if the chemical potential of matter at $\beta$-equilibrium reaches a large enough value, particles with non-zero strangeness can be stabilized in the system. In particular one can expect by charge neutrality arguments that $\Sigma^-$ and $\Lambda$ hyperons could be the first species to appear. The larger mass of these hadrons ($m_\Sigma\simeq1193$~MeV and $m_\Lambda\simeq1116$~MeV), together with the fact that they become distinguishable with respect to nucleons, lowers the energy of the system, making the EoS softer. Several calculations performed in the past using the available hyperon-nucleon interactions all confirmed this picture (see for example Refs.~\cite{Dapo:2010,Schulze:2011,Vidana:2011,Massot:2012}). However, the softening of the EoS is such that the predicted maximum mass of a NS is not compatible with the current observational data. 

This apparent puzzle could be solved if the interaction between hyperons and nucleons becomes so repulsive at large baryon density, that the system does not find energetically convenient to push the creation of hyperons beyond a certain limit, thereby preventing an excessive softening of the EoS. In this spirit, the aim of this work was to perform a deeper analysis of the hyperon-nucleon interaction starting from the data available for $\Lambda$~hypernuclei. Our scheme is based on the use of Quantum Monte Carlo methods (more specifically the Auxiliary Field Diffusion Monte Carlo method~\cite{Schmidt:1999,Gandolfi:2007,Gandolfi:2009}) in order to solve for the ground state of an Hamiltonian including a realistic local potential which contains hyperon-nucleon and hyperon-nucleon-nucleon terms. The solutions allow us for discussing the systematics of the binding energy of the hyperon in hypernuclei, and consequently fine tuning the interaction itself. The final outcome of these calculations should be a realistic potential to be used in determining the properties of hyperon-nucleon matter, and in this paper some of the progress along this way will be presented.

\section{$\Lambda N$ Potential}
In this paper we will consider only systems including neutral $\Lambda$ hyperons. So far, experimental binding energies and excitation energies are available only for a limited set of $\Lambda$~hypernuclei~\cite{Juric:1973,Cantwell:1974,Pile:1991,Hashimoto:2006,Nakamura:2013}, which is hopefully going to be extended by ongoing measurements at several facilities worldwide. The number of available $p\Lambda$ scattering events is also relatively small~\cite{deSwart:1971,Kadyk:1971,Ahn:2005}, but enough to constrain the main contributions of the hyperon-nucleon interaction.
Bodmer, Usmani~\emph{et al.} proposed a phenomenological potential inspired by the Argonne-Urbana forces (see~\cite{Bodmer:1988,Usmani:1995,Usmani:2006,Usmani:2008} and references therein), that is sufficient to capture the physical information that is presently at disposal. 

\begin{figure}[ht]
	\centering
		\subfigure[\label{fig:LN}]%
			{\includegraphics[height=3.2cm]{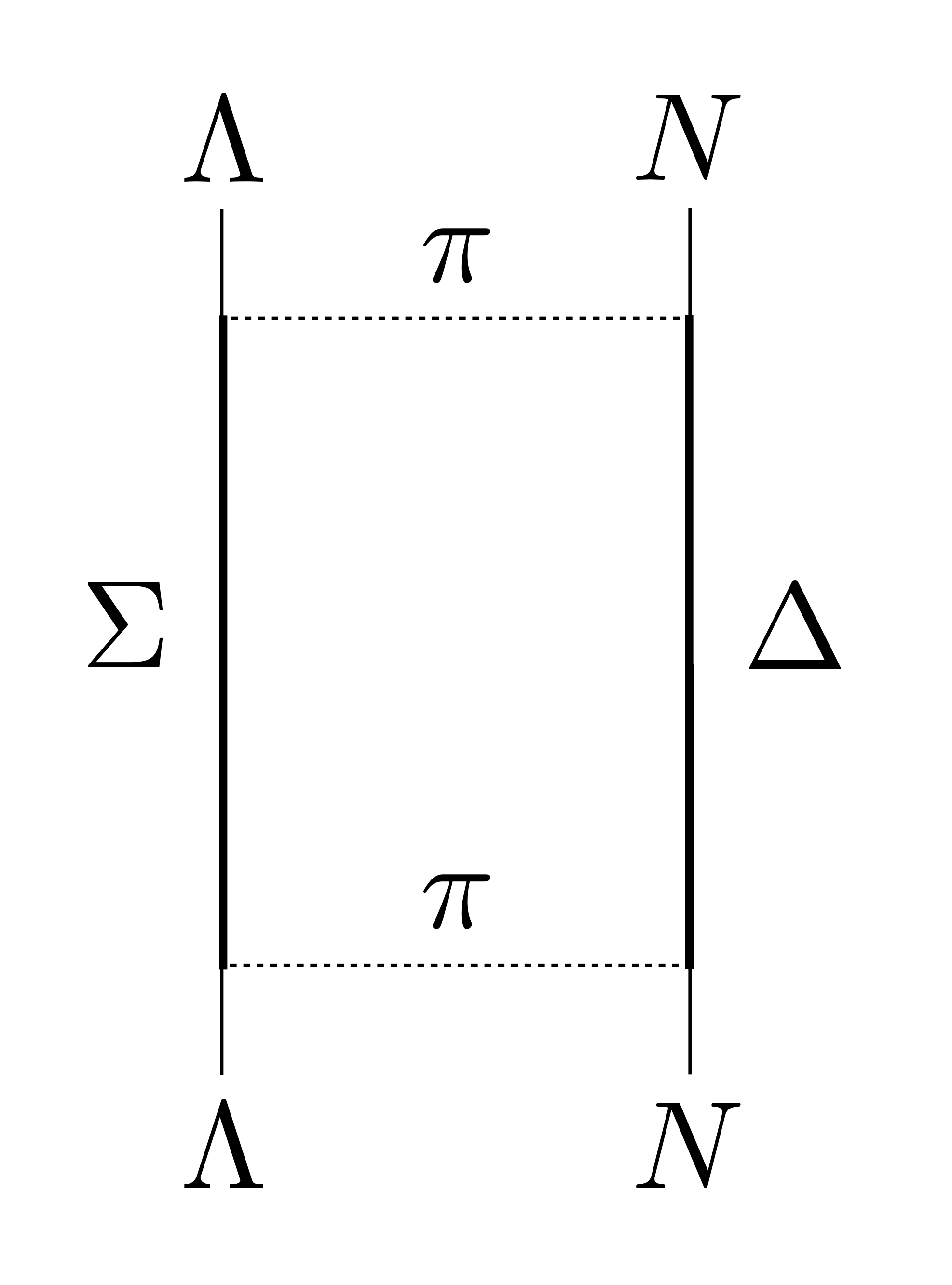}}
		\goodgap\goodgap\goodgap
		\subfigure[\label{fig:LNN_sw}]%
			{\includegraphics[height=3.2cm]{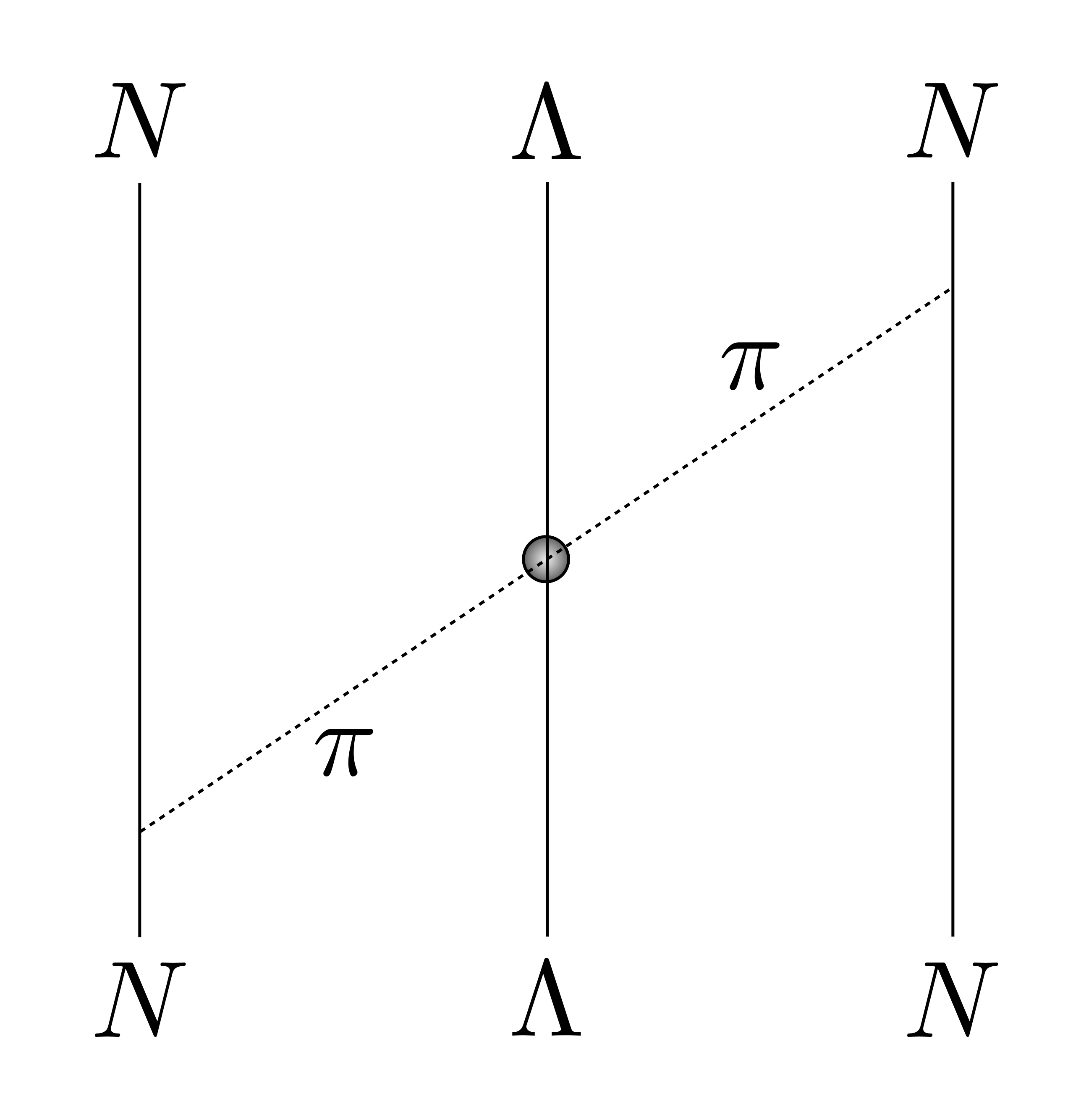}}
		\goodgap\goodgap\goodgap
		\subfigure[\label{fig:LNN_pw}]%
			{\includegraphics[height=3.2cm]{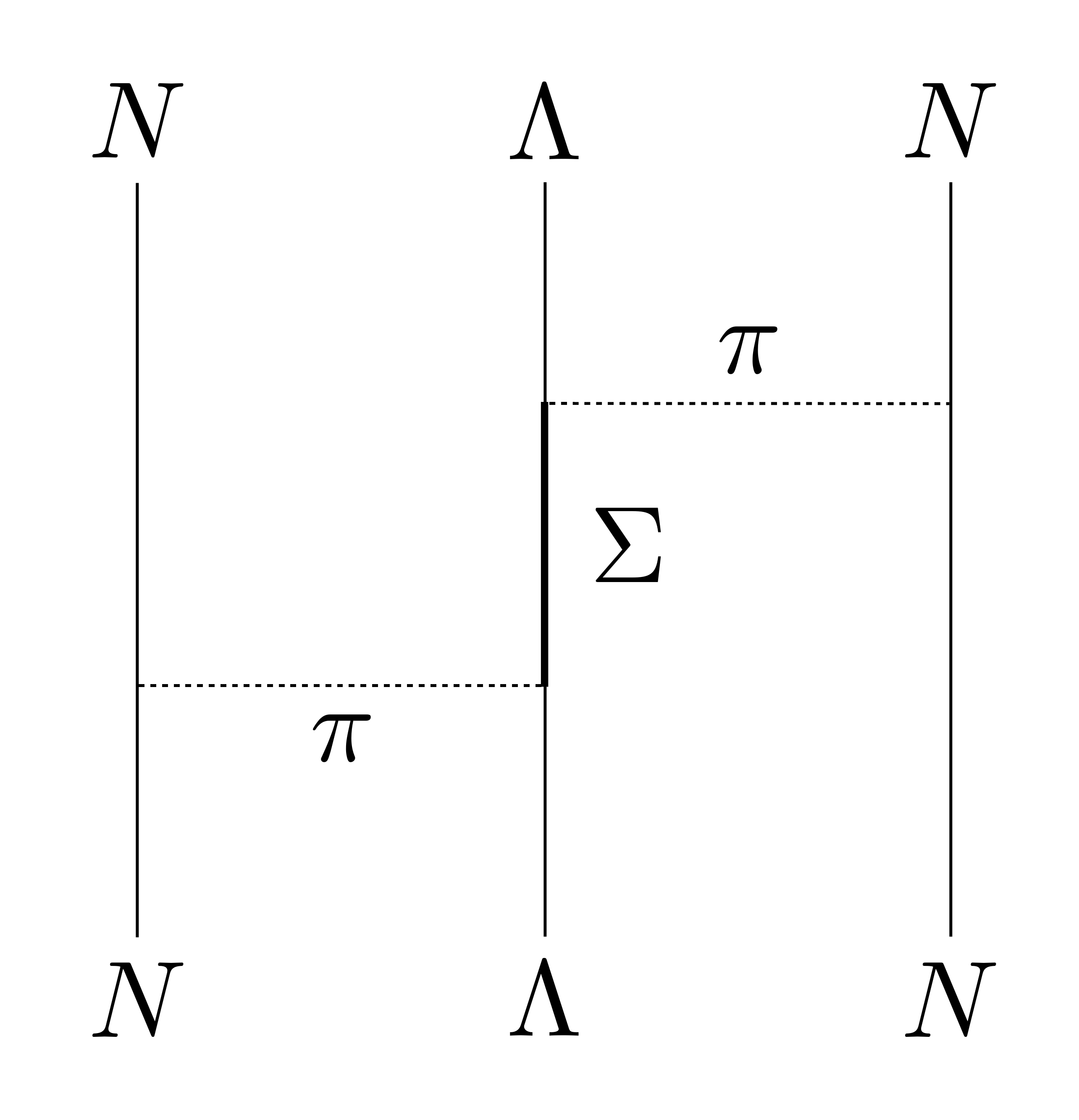}}
		\goodgap\goodgap\goodgap
		\subfigure[\label{fig:LNN_d}]%
			{\includegraphics[height=3.2cm]{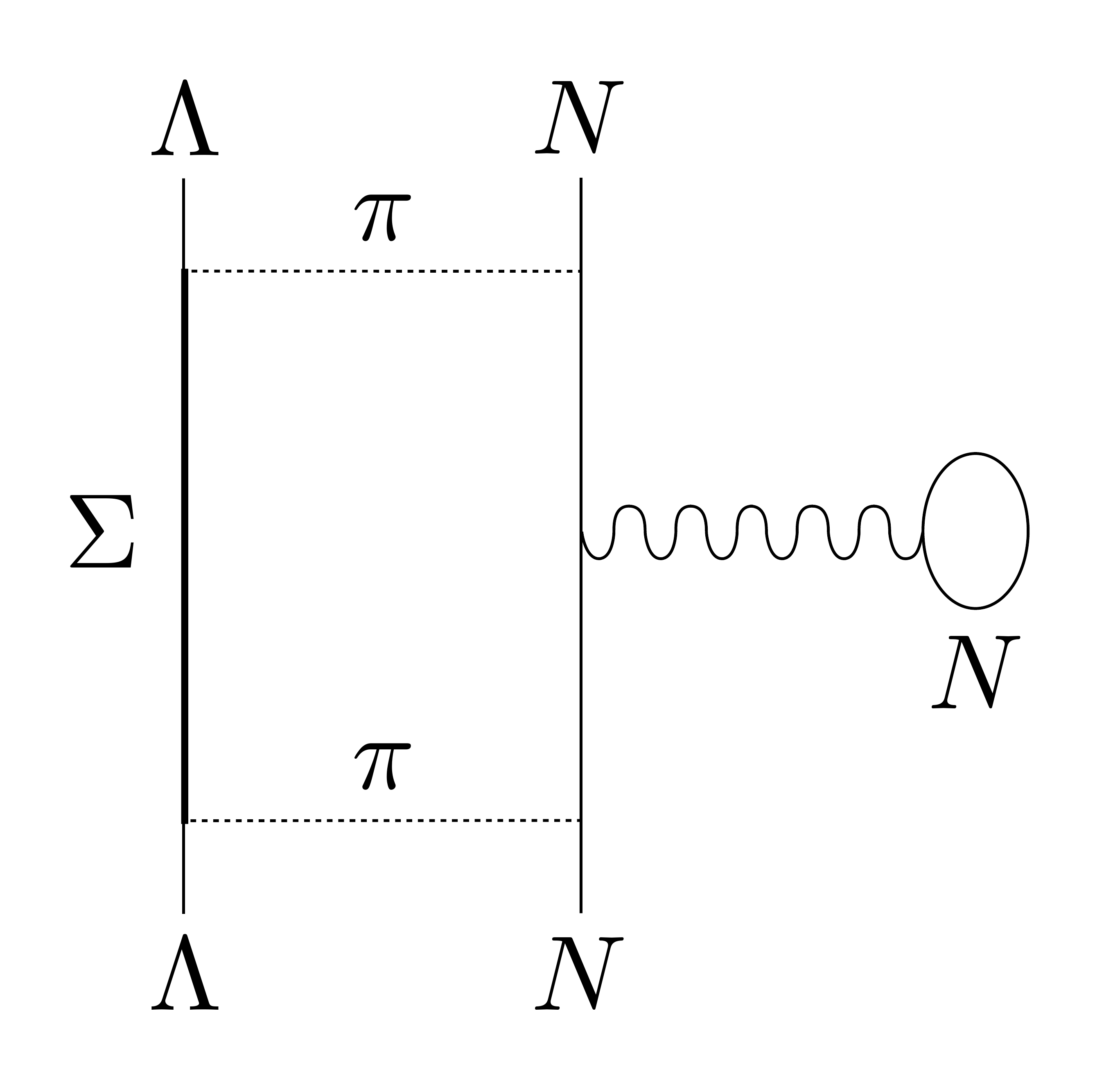}}
	\caption[]{A schematic representation of the phenomenological $\Lambda$-nucleon interaction. 
		In this scheme $2\pi$-exchange processes between nucleons and the
		$\Lambda$ particle appear as two-body ($a$) and three-body (($b$), ($c$) and ($d$)) contributions. 
		The last three will be labelled in the following as $S$-wave, $P$-wave and Dispersive terms respectively.}
	\label{fig:LN-LNN} 
\end{figure}

It should be noticed that isospin conservation prevents the occurrence of vertices $\Lambda\pi\Lambda$. As a consequence, the $\Lambda$-nucleon interaction requires at least the exchange of two pions. In Fig.~\ref{fig:LN-LNN} we list all the possible occurring processes of this kind. As it can be seen, only one of this contributions involve two particles. All the others are three-body forces of the $\Lambda NN$ kind. It should be stressed out that this specific subdivision in two and three-body terms is somewhat depending on the scheme used. In the $\Lambda N$ channel we should also include a process involving the exchange of kaons, which would give rise to a permutation of $\Lambda$ and nucleons. For technical reasons we will include this contribution effectively in the coefficients of the interaction. In Refs.~\cite{Usmani:2006,Usmani:2008} Usmani~\emph{et al.} reported this term to contribute order 10\% and 30\% of the central interaction.

The two-body terms has the form:
\begin{align}
	v_{\lambda i}=v_{0}(r_{\lambda i})(1-\varepsilon+\varepsilon\,\mathcal P_x)
	+\frac{1}{4}v_\sigma T^2_\pi(r_{\lambda i})\,{\bm\sigma}_\lambda\cdot{\bm\sigma}_i \;,
	\label{eq:V_LN}
\end{align}
where $v_0(r)=v_c(r)-\bar v\,T_{\pi}^{2}(r)$ is a central term. The coefficients
$\bar v=(v_s+3v_t)/4$ and $v_\sigma=v_s-v_t$ are the spin-average and
spin-dependent strengths, where $v_s$ and $v_t$ denote singlet- and
triplet-state strengths, respectively.  Both the spin-dependent
and the central radial terms contain the usual regularized OPE tensor
operator $T_\pi(r)$
\begin{align}
	T_{\pi}(r)=\left[1+ \frac{3}{\mu_\pi r}+ \frac{3}{(\mu_\pi r)^2} \right]
	\frac{\e^{-\mu_\pi r}}{\mu_\pi r}\Bigl(1-\e^{-cr^2}\Bigr)^2\;,\label{eq:T_pi}
\end{align}
where $\mu_\pi$ is the pion reduced mass
\begin{align}
	\mu_\pi=\frac{1}{\hbar}\frac{m_{\pi^0}+2\,m_{\pi^\pm}}{3}\quad\quad\frac{1}{\mu_\pi}\simeq 1.4~\text{fm} \;.
\end{align}
The three-body terms have the following expressions:
\begin{align}
	v_{\lambda ij}^{S}&=
		C_S\,Z\left(r_{\lambda i}\right)Z\left(r_{\lambda j}\right)\,
		{\bm\sigma}_{i}\cdot\hat{\bm r}_{i\lambda}\,
		{\bm\sigma}_{j}\cdot\hat{\bm r}_{j\lambda}\,{\bm\tau}_{i}\cdot{\bm\tau}_{j}\;,\label{eq:V_LNN_S}\\[0.7em]
	v_{\lambda ij}^{P}&=-\frac{C_P}{6}
		\Bigl\{X_{i\lambda}\,,X_{\lambda j}\Bigr\}\,{\bm\tau}_{i}\cdot{\bm\tau}_{j}\;,\label{eq:V_LNN_P}\\[0.7em]
	v_{\lambda ij}^{D}&=W_D\,
		T_{\pi}^{2}\left(r_{\lambda i}\right)T^{2}_{\pi}\left(r_{\lambda j}\right)
		\!\!\bigg[1+\frac{1}{6}{\bm\sigma}_\lambda\!\cdot\!\left({\bm\sigma}_{i}+{\bm\sigma}_{j}\right)\bigg]\;.\label{eq:V_LNN_D}
\end{align}
The functions $X_{\lambda i}$ and $Z(r)$ are defined by
\begin{align}
	X_{\lambda i}&=Y_{\pi}(r_{\lambda i})\;\bm\sigma_{\lambda}\cdot\bm\sigma_{i}+
	T_{\pi}(r_{\lambda i})\;S_{\lambda i}\;,\\[0.7em]
	Z(r)&=\frac{\mu_\pi r}{3} \Bigl[Y_\pi(r)-T_\pi(r)\Bigr]\;,
\end{align}
where the function $Y_\pi(r)$ is the usual regularized Yukawa potential
\begin{align}
	Y_\pi(r)=\frac{\e^{-\mu_\pi r}}{\mu_\pi r}\Bigl(1-\e^{-cr^2}\Bigr)\;,
	\label{eq:Y_pi}
\end{align}
and $S_{\lambda i}$ is the tensor operator.

The parameters appearing in Eq.~(\ref{eq:V_LN}) have been determined by fitting the available scattering data, and have never been varied in our studies. Details on the parametrization can be found in Refs.~\cite{Usmani:1995,Usmani:2006,Usmani:2008}. For a complete review see Ref.~\cite{Lonardoni:2013_thesis}.

\section{Computational Methods}
We model hypernuclei made up of $A-1$ nucleons and one $\Lambda$ by the non-relativistic Hamiltonian
\begin{align}
	\hat{H}=\sum_{i=1}^{A-1}\frac{\hat{p_i}^2}{2m_N}+\frac{\hat{p}_{\Lambda}^2}{2m_\Lambda}+\sum_{i<j}v_{ij}(r_{ij})
	+\sum_{i=1}^{A-1} v_{\lambda i}(r_{\Lambda i})+\sum_{i<j}v_{\lambda ij}(r_{\Lambda i},r_{\Lambda j})\;,
\end{align}
where $v_{ij}$ is the two-nucleons interaction.
In order to obtain information about the ground state properties we use a stochastic projection algorithm, namely the Auxiliary Field Diffusion Monte Carlo (AFDMC) method~\cite{Schmidt:1999,Gandolfi:2007,Gandolfi:2009}. AFDMC is based on propagating a number of points in the extended configuration space (including the coordinates, spin and isospin degrees of freedom), by sampling an approximation of the Green's function of the imaginary time Hamiltonian operator
\begin{align}
	\Psi(R,S,T;\tau+d\tau)=\sum_{S',T'}\int dR'\,G(R,S,T;R',S',T';d\tau)\,\Psi(R',S',T';\tau)\;,
\end{align}
where $R=\{{\bm r}_1\dots{\bm r}_A\}$, $S=\{{\bm \sigma}_1\dots{\bm \sigma}_A\}$, $T=\{{\bm \tau}_1\dots{\bm \tau}_A\}$ are the coordinates, spin and isospin of the baryons respectively. The sum over $S'$ and $T'$ has to be intended as the sum over all the possible spin/isospin states of the baryons.
The approximated Green's function for a small step $d\tau$ in imaginary time is then written as:
\begin{align}
	G(R,S,T;R',S',T';d\tau)\simeq\left(\frac{m}{2\pi\hbar^2d\tau}\right)^{\frac{3A}{2}}\e^{-\frac{m(R-R')}{2\hbar^2d\tau}}\e^{-\left[\hat V(R,S,T)-E\right]d\tau}\;.
\end{align}
In general the last factor contains terms that are operatorial and/or non-local. In standard Quantum Monte Carlo calculations one is usually limited to use in the Green's function interactions that are local and at most quadratic in the spin/isospin operators, with the only exception of spin-orbit terms~\cite{Pieper:2008}. Therefore, the sum over spin and isospin states requires the use of a wave function that has a number of components exponentially growing with $A$. The AFDMC method makes use of the Hubbard-Stratonovich transformation to make only single particle spin/isospin operators in the Green's function. The advantage is the possibility to work on a single particle representation of the wave function, thereby reducing the computational cost of the calculation from exponential to cubic in $A$. On the other hand, this method makes more challenging the use of accurate wave functions including quantum correlations in operatorial channels. This has the consequence of reducing the accuracy that is achievable when imposing the approximate sampling constraints necessary to avoid the exponential growth of the variance due to the Fermion Sign Problem.
Because of such technical reasons we limited ourselves to a nucleon-nucleon interaction that can be considered under control in AFDMC calculations, namely the Argonne V4' potential~\cite{Wiringa:2002}. This force is known to give a reasonable binding energy for small nuclei, but to strongly overbind medium/heavy nuclei. However, we are interested in making a connection with experiments only through the $\Lambda$~separation energy defined as
\begin{align}
	B_\Lambda=B_{\text{nuc}}(A-1)-B_{\text{hyp}}(A-1,\Lambda)\;,
\end{align}
where $B_{\text{nuc}}$ and $B_{\text{hyp}}$ are the total binding energies of the the nucleus and the $\Lambda$-hypernucleus with $A-1$ nucleons respectively. The hypothesis we make, rather well justified a posteriori, is that in the difference the contributions due to the nucleon-nucleon interaction mostly cancel out. Besides the reasonable agreement we obtain for the available experimental results for medium mass-hypernuclei, this assumption is supported by the fact that changing the interaction from AV4' to AV6'~\cite{Wiringa:2002} (which gives a systematic underbinding of nuclei) does not seem to give a strong influence on the results~\cite{Lonardoni:2013_PRC(R)}. The presence of the $\Lambda$ hyperon also requires a modification of the wave function and a special treatment of the kinetic energy, in order to correctly subtract the contributions of the center of mass motion. Details can be found in Refs.~\cite{Lonardoni:2013_PRC(R),Lonardoni:2013_HYP2012,Lonardoni:2013_thesis,Lonardoni:2013_PRC}.

\section{Results}

\subsection{Hypernuclei}

As previously mentioned, the key point of this work is to begin the journey towards an accurate phenomenological hyperon-nucleon interaction that can be employed in many-body Quantum Monte Carlo calculations. Given that at present the number and quality of available scattering data do not allow for a substantial improvement of the $\Lambda N$ contribution, we focused on refitting the coefficients appearing in the three-body $\Lambda NN$ term. 
It emerges form the calculations that the binding energies are substantially insensitive to the value of the coefficient of the $S$-wave term in the $2\pi$-exchange channel $C_S$, that was fixed to the value $1.5$~MeV~\cite{Usmani:2008}. We are therefore left with two parameters to determine. This can be done by reproducing the experimental results in two hypernuclei. We chose two closed-shell systems, which we believe to be much less prone to suffer from the necessary approximations related to the use of AFDMC, namely $^5_\Lambda$He and $^{17}_{~\Lambda}$O.

\begin{figure}
	\centering
	\includegraphics[scale=0.5]{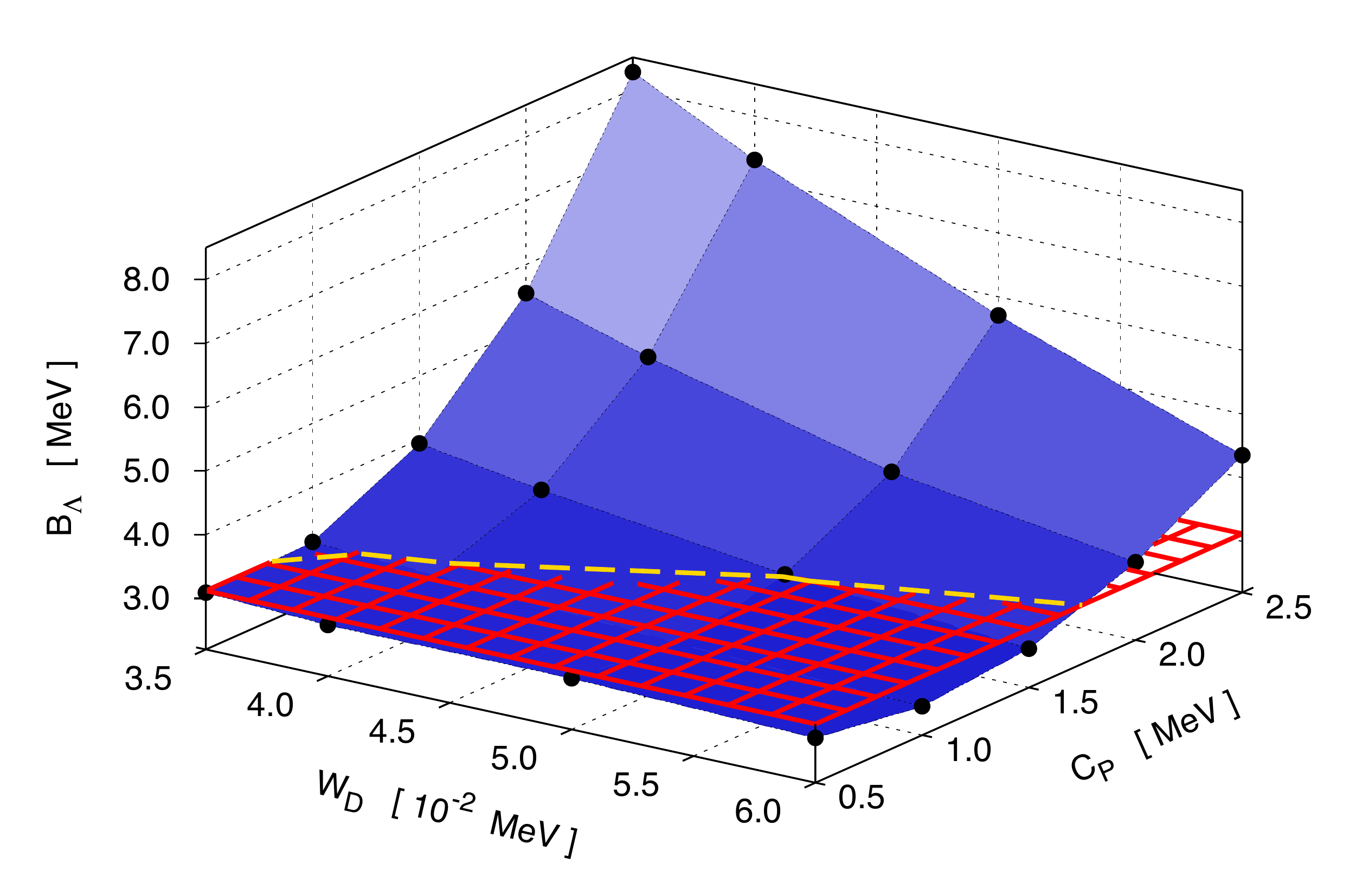}
	\caption{The $\Lambda$~separation energy $B_\Lambda$ as a function of the coefficients $W_D$ and $C_P$ of the $\Lambda NN$ force, 
		relative to the the $P$-wave component of the $2\pi$-exchange term and the dispersive term respectively. 
		The red plane is the experimental value from Ref.~\cite{Juric:1973}.}
	\label{fig:Wd-Cp}
\end{figure}

In Fig.~\ref{fig:Wd-Cp} the dependence of $B_\Lambda$ in $^5_\Lambda$He as a function of the parameters $W_D$ and $C_P$ is shown. The parametrizations compatible with the experimental result $B_\Lambda=3.12(2)$~MeV~\cite{Juric:1973} (represented by the red plane) are those found at the intersection of the two surfaces. Among them, we picked the one that better reproduces the (extrapolated) $B_\Lambda$ in $^{17}_{~\Lambda}$O~\cite{Usmani:1995}. The resulting values are $W_D=0.035$~MeV and $C_P=1.00$~MeV~\cite{Lonardoni:2013_PRC}. With this parametrization, we proceeded to compute the $\Lambda$~separation energy for a number of other light and medium-heavy hypernuclei. In particular, we were interested in testing the saturation property of $B_\Lambda$, which is of extreme importance in view of extending the calculations to infinite matter.
 
\begin{figure}
	\centering
	\includegraphics[scale=0.5]{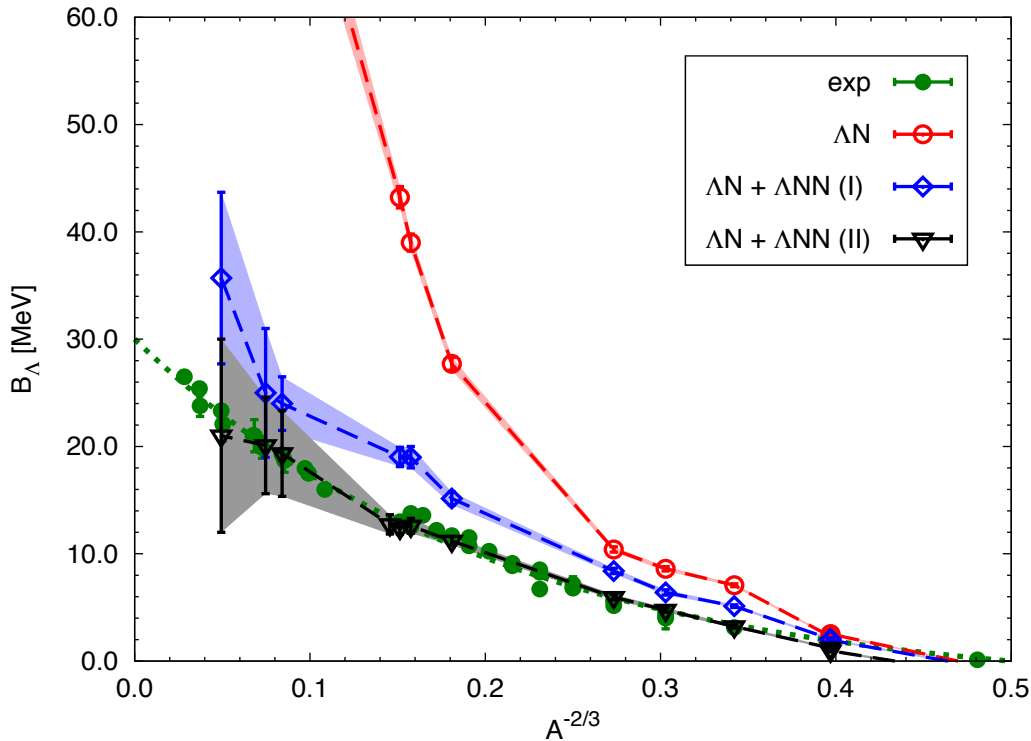}
	\caption{The $\Lambda$~separation energy $B_\Lambda$ as a function of $A^{-2/3}$. 
		The red dots are the results obtained with an Hamiltonian containing a two-body $\Lambda N$ force only. 
		Blue diamonds are AFDMC calculations with the original 2- and 3-body $\Lambda N$ potential by Usmani~\cite{Usmani:1995_3B}. 
		Black triangles are the current AFDMC results with the refitted 3-body interaction~\cite{Lonardoni:2013_PRC}. 
		Green dots are experimental results. Lines and bands are drawn as a guid to the eye.}
	\label{fig:BL}
\end{figure}

Given the available computational resources we were able to simulate closed-shell nucleus/hypernucleus pairs up to $^{91}_{~\Lambda}$Zr~\cite{Lonardoni:2013_PRC(R),Lonardoni:2013_PRC}. The results are summarized in Fig.~\ref{fig:BL}, where $B_\Lambda$ is plotted as a function of $A^{-2/3}$. As it can be seen, the effect of including the $\Lambda NN$ term in the Hamiltonian is very strong. It provides the repulsion necessary to realistically reproduce the limiting value of $B_\Lambda$. One important test of our calculation also comes from the fact that the new parametrization of the $\Lambda NN$ term, that was fixed only using data for two hypernuclei, is very good in reproducing the overall behavior of the experimental data. Some discrepancies appear at very low masses ($A<5$), where other effects should be expected, such as a rather strong contribution from charge symmetry breaking components of the interaction.

\begin{figure}
	\centering
	\includegraphics[scale=0.5]{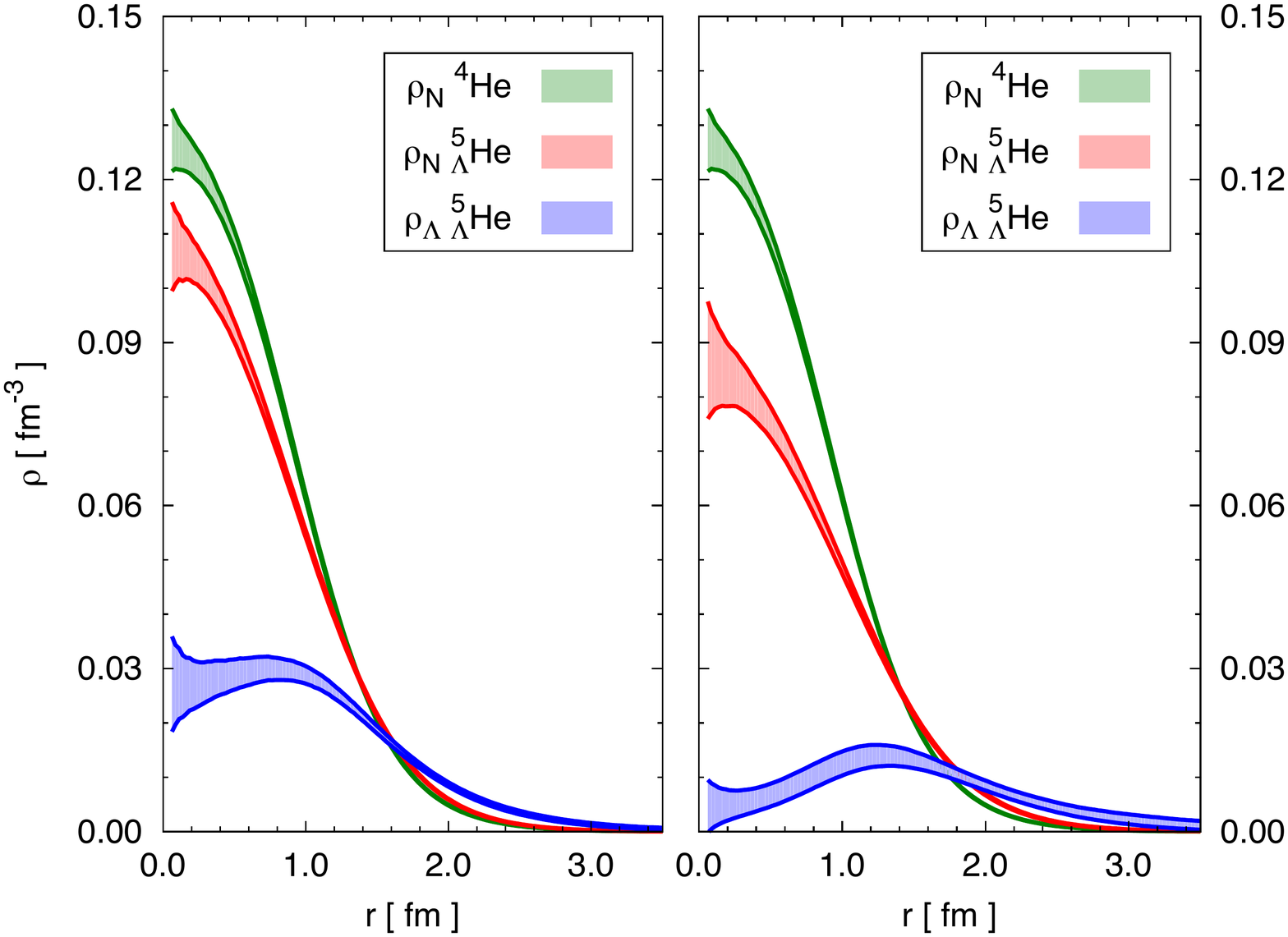}
	\caption{Single particle densities for nucleons in $^4$He [green, upper banded curve] 
		and for nucleons [red, middle banded curve] and
        the lambda particle [blue, lower banded curve] in $^5_\Lambda$He.
        In the left panel the results for the two-body $\Lambda N$
        interaction alone.  In the right panel the results with the inclusion
        also of the three-body hyperon-nucleon force with the new set of parameters.}
	\label{fig:rho_He5L}
\end{figure}

In Fig.~\ref{fig:rho_He5L} we report the results for the single particle
densities for $^4$He and $^5_\Lambda$He. Densities here are computed without taking into 
account the effect of the displacement of the center of mass due to the difference between $m_\Lambda$ and $m_N$. 
The green curves are the densities of nucleons in the nucleus, while the red and
blue curves are, respectively, the density of nucleons and of the
lambda particle in the hypernucleus. In the left panel the results are
obtained using AV4' for the nuclear part and the two-body $\Lambda N$
interaction alone for the hypernuclear component. In the right panel
the densities are calculated with the full two- plus three-body 
hyperon-nucleon interaction in the new parametrization. 

The correct
estimators for the single particle densities are obtained starting from
the mixed DMC results and the variational ones via the positive defined
relation~\cite{Pieper:2008}:
\begin{align}
        \langle\mathcal O\rangle_{real}=
        \frac{\langle\psi_0|\mathcal O|\psi_0\rangle}{\langle\psi_0|\psi_0\rangle}
        =\frac{\left(\frac{\langle\psi_T|\mathcal O|\psi_0\rangle}{\langle\psi_T|\psi_0\rangle}\right)^2}
        {\frac{\langle\psi_T|\mathcal O|\psi_T\rangle}{\langle\psi_T|\psi_T\rangle}}
        =\frac{\langle\mathcal O\rangle_{\scriptscriptstyle{DMC}}^2}{\langle\mathcal O\rangle_{\scriptscriptstyle{VMC}}}\;,
\end{align}
where $\mathcal O$ is the density operator
$\hat\rho=\sum_{i}\delta(r-r_i)$, $\psi_T$ is the trial wave function and $\psi_0$ is the projected ground state wave function.
The addition of the $\Lambda$ particle to the nuclear core of $^4$He has
the effect to reduce the nucleon density in the center. The $\Lambda$~particle tries to localize close to $r=0$, enlarging therefore the
nucleon distribution. When the three-body $\Lambda NN$ interaction
is turned on (right panel of Fig.~\ref{fig:rho_He5L}), the repulsion
moves the nucleons to large distances but the main effect is that the
hyperon is pushed away from the center of the system. 

\begin{figure}
	\centering
	\includegraphics[scale=0.9]{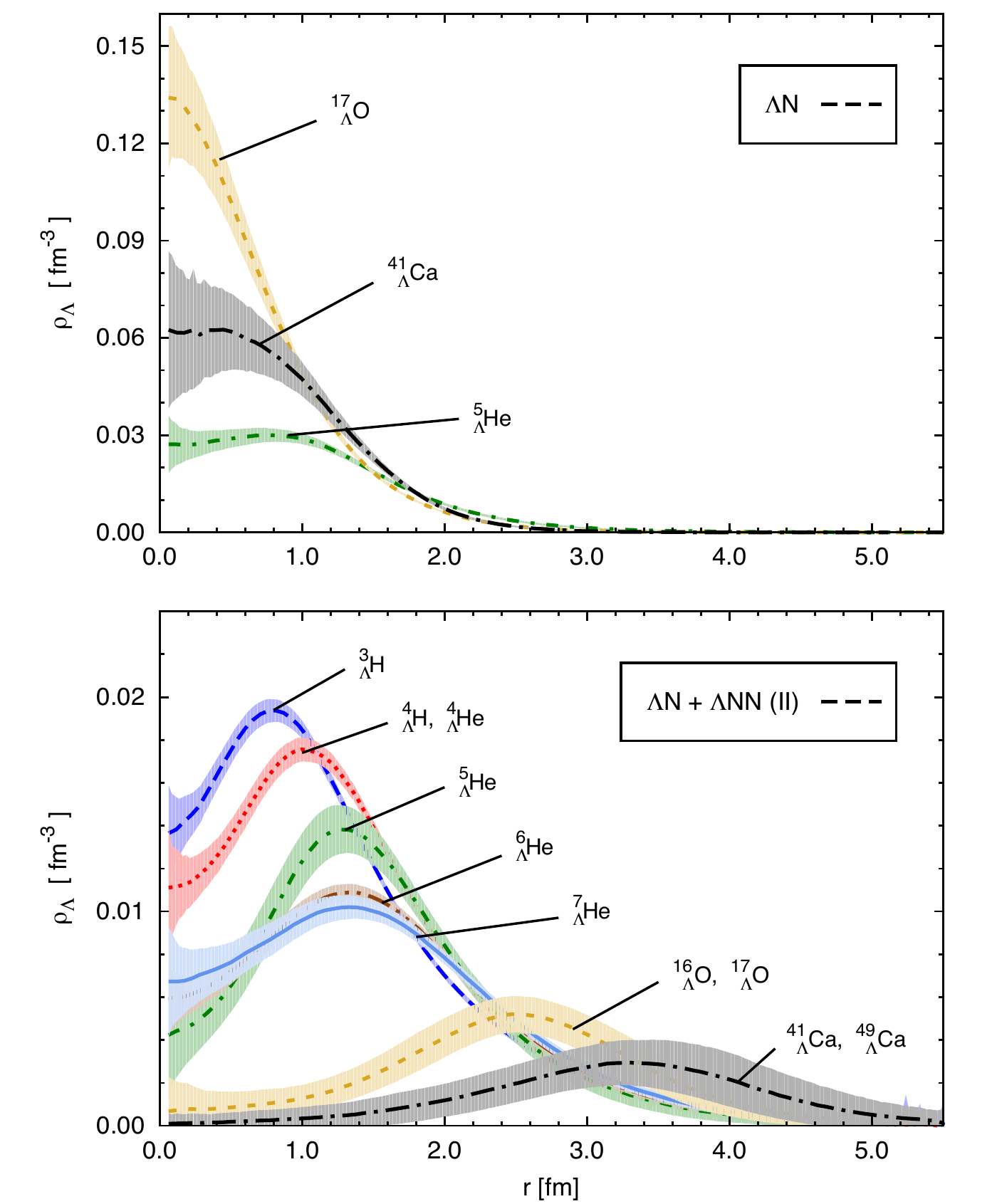}
	\caption{Single particle densities for the $\Lambda$ particle in different hypernuclei. 
		Top panel reports the results for the two-body $\Lambda N$ interaction alone.  
		Bottom panel shows the results when the three-body hyperon-nucleon interaction with the new set of
		parameters is also included.}
	\label{fig:rho_L}
\end{figure}

As can be seen from Fig.~\ref{fig:rho_L}, this effect is much more evident for large $A$. When
the hypernucleus is described by the $\Lambda N$ interaction alone, the
$\Lambda$ particle is localized near the center, in the range $r<2$~fm
(top panel of Fig.~\ref{fig:rho_L}). The inclusion of the three-body
$\Lambda NN$ potential forces the hyperon to move from the center, in a
region that roughly correspond to the skin of nucleons. It should be noticed that 
the nuclear densities given by the AV4' are widely overestimated. However we want
to point out a couple of interesting facts. First of all, when using the AV6' potential we found the same effects on the $\Lambda$ particle, confirming the importance of the three-body hyperon-nucleon interaction and its strongly repulsive nature.
Second, it can be noticed that the position of the peak of the $\Lambda$ density roughly correspond to the distance from the center at which the nuclear density is about the saturation value. This might indicate that when using a correct $NN$ force the hyperon could be found more towards the center. This might also reconcile the observed shrinkage effect of the core nucleus~\cite{Hashimoto:2006,Tanida:2001}, as obtained for example in Ref.~\cite{Sinha:2002}.

\subsection{Infinite Matter}
The prediction of the structure of a NS requires the knowledge of the EoS of an infinite homogeneous matter. In the NS core weak interactions always sustain the presence of a finite fraction of protons, electrons and muons. Chemical equilibrium among these species determines the composition of the NS core. As already mentioned, when the density increases hyperons can appear. In general, it is required that the number of baryons and the charge neutrality are preserved. 

\begin{figure}
	\centering
	\includegraphics[scale=0.5]{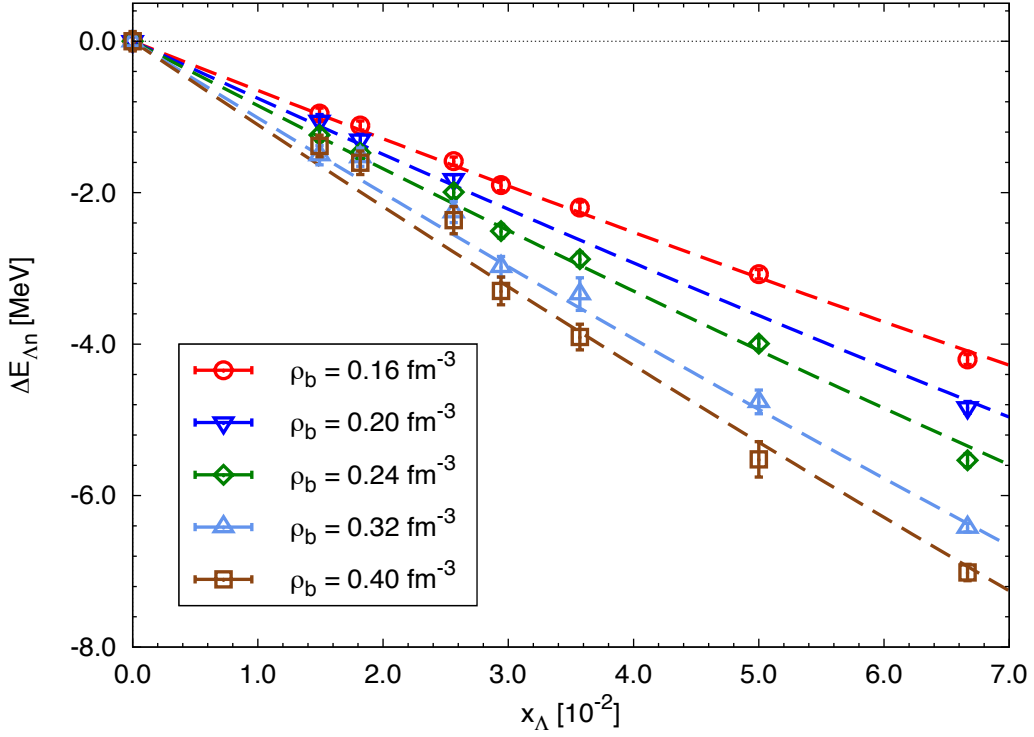}
	\caption{The energy difference $\Delta E_{\Lambda n} = E_{\Lambda n}-E_{PNM}$ per baryon as a 
		function of the $\Lambda$~fraction $x_\Lambda$ in a mixed $\Lambda$-neutron matter for different 
		values of the total baryon density $\rho_b$. The points are AFDMC results, while the corresponding 
		lines are linear fits according to Eq.~(\ref{eq:EoS}).}
	\label{fig:E_x}
\end{figure}

In order to simplify the description, as a first approximation, the interior of a NS can be assumed to be made of neutrons only. In this case the chemical equilibrium between hyperons and neutrons is simply given by the condition $\mu_\Lambda=\mu_n$. In principle it is possible to compute the chemical potential of the species $\kappa$ as:
\begin{align}
	\mu_\kappa=\frac{\partial \mathcal E\left(\{\rho_\kappa\}\right)}{\partial \rho_\kappa}\Bigg|_V\;,
\end{align}
where $\mathcal E$ is the total energy density (time component of the relativistic energy-momentum density 4-vector) of the system. Besides of the masses of the baryons, $\mathcal E$ must include the interaction energy. In the limit of a small fraction of hyperons present in the neutron medium, it is sensible to express the $\Lambda N$ interaction energy per baryon as a function of the total baryon density $\rho_b$ in the form:
\begin{align}
	E_{\Lambda n}(\rho_b,x_\Lambda)=E_{PNM}(\rho_b)-S_{\Lambda n}(\rho_b)\,x_\Lambda\;,
	\label{eq:EoS}
\end{align}
where $E_{PNM}$ is the energy per baryon of pure neutron matter (PNM), $x_\Lambda$ is the hyperon fraction $x_\Lambda=\rho_\Lambda/\rho_b$ and $S_{\Lambda N}(\rho_b)$ could be considered as an analog of the symmetry energy. In the previous equation we neglect terms order $x_\Lambda^2$ which might be in principle non negligible. We should remember that being $m_\Lambda \ne m_n$, there is no reason for linear terms in the hyperon asymmetry not to be present in the EoS. 
In Fig.~\ref{fig:E_x} we report the difference between the energy of $\Lambda$-neutron matter for finite hyperon fractions and the energy of PNM as a function of $x_\Lambda$, and for different total baryon densities. These results are obtained by simulating a system in which we add one or two hyperons in a periodic cubic box containing a given number of neutrons at a given density. Particular care must be paid here to correct for finite size effects. As it can be seen from the figure, the behavior of the energy difference can be reproduced by a linear fit. 

\begin{figure}
	\centering
	\includegraphics[scale=0.5]{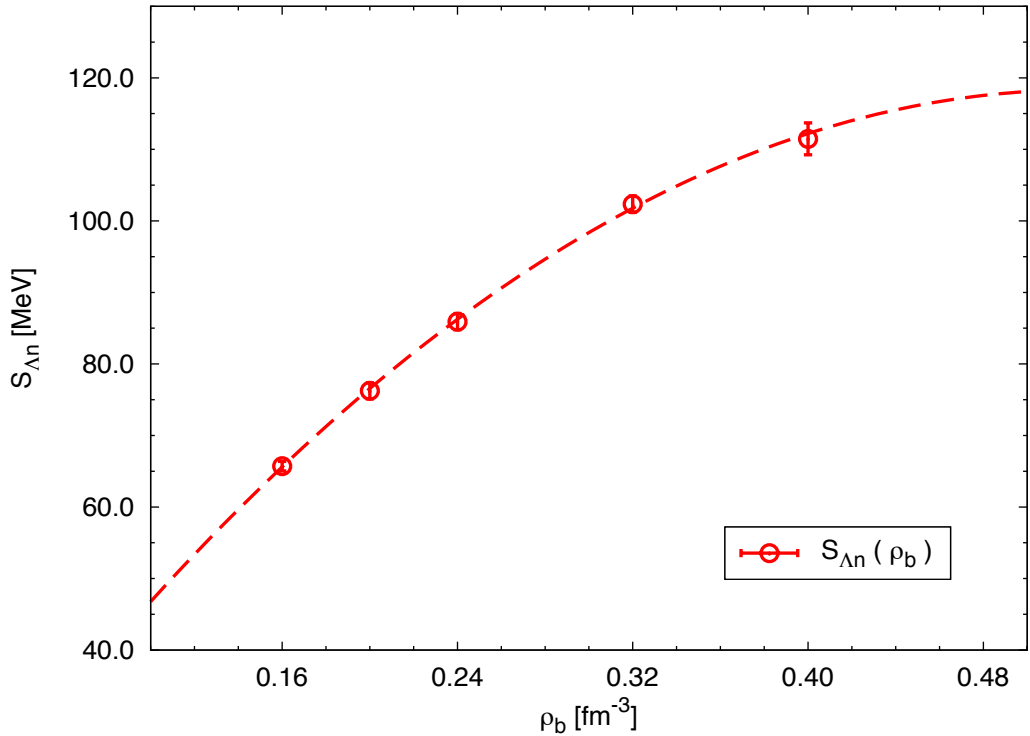}
	\caption{The coefficient $S_{\Lambda n}$ (analog of the symmetry energy in a mixed $\Lambda$-neutron matter) 
		as a function of the total baryon density $\rho_b$. The dots are the values obtained by fitting the AFDMC results 
		for the energy difference defined from Eq.~(\ref{eq:EoS}). The dashed line is a fit to the data.} 
	\label{fig:S_rho}
\end{figure}

From the slopes of the linear fits it is possible to obtain the behavior of $S_{\Lambda n}(\rho_b)$, which is reported in Fig.~\ref{fig:S_rho}.
By knowing the $S_{\Lambda n}$ function, the EoS of the $\Lambda$-neutron matter is given by Eq.~(\ref{eq:EoS}) as a function of both $\rho_b$ and $x_\Lambda$. It is then possible to impose the chemical equilibrium condition between hyperons and neutrons and derive the threshold density for the appearance of hyperons and the $\Lambda$ fraction as a function of the total baryon density. These are key ingredients in the description of the $\Lambda$-neutron medium that affect the derivation of the mass-radius relation and the maximum mass of a NS. First steps in this direction have been taken and preliminary results are reported in Ref.~\cite{Lonardoni:2013_thesis}.

\section{Conclusions}
Ground state properties of hypernuclei over a wide range of masses were studied by means of the Auxiliary Field Diffusion Monte Carlo method. The main outcome of this is the determination of a realistic three-body local $\Lambda NN$ interaction that reproduces with high accuracy the saturation properties of the hyperon separation energy. We also showed how by this method a good insight on the properties of homogeneous matter can be gained. In particular, we determined the behavior of $S_{\Lambda n}$, an analog of the symmetry energy for a mixed hyperon-neutron matter, which is the necessary ingredient to determine the equation of state of $\Lambda$-neutron matter, of great interest for the determination of the inner structure of neutron stars.

\ack
This work has been partially performed at LISC, Interdisciplinary
Laboratory for Computational Science, a joint venture of the University
of Trento and Bruno Kessler Foundation.  Support and computer time
were partly made available by the AuroraScience project (funded by the
Autonomous Province of Trento and INFN), and by Los Alamos Open Supercomputing.
This research used also resources of the National Energy Research
Scientific Computing Center, which is supported by the Office of
Science of the U.S. Department of Energy under Contract No.
DE-AC02-05CH11231.
The work of S.~G. was supported by the Department of Energy Nuclear
Physics Office, by the NUCLEI SciDAC program, and by a Los Alamos LDRD
early career grant.

%\bibstyle{iopart-num}
%\bibliography{biblio}
\providecommand{\newblock}{}
\expandafter\ifx\csname url\endcsname\relax
  \def\url#1{{\tt #1}}\fi
\expandafter\ifx\csname urlprefix\endcsname\relax\def\urlprefix{URL }\fi
\providecommand{\eprint}[2][]{\url{#2}}

\end{document}